\documentclass[11pt]{article}

\usepackage[margin=1in]{geometry}

\usepackage[T1]{fontenc}
\usepackage[utf8]{inputenc}
\usepackage{microtype}
\usepackage{newtxtext,newtxmath} 

\usepackage{xcolor}
\usepackage{url}

\usepackage{amsmath,amsfonts}
\usepackage{amssymb}

\usepackage{graphicx}
\usepackage{array}
\usepackage{booktabs}
\usepackage{multirow}
\usepackage{tabularx}
\usepackage{longtable}
\usepackage{subcaption}
\usepackage{ragged2e}
\usepackage{placeins}  

\newcolumntype{L}[1]{>{\raggedright\arraybackslash}p{#1}}
\newcolumntype{M}[1]{>{\raggedright\arraybackslash}m{#1}}

\usepackage{algorithm}
\usepackage{algpseudocode}

\usepackage[colorlinks=true,allcolors=blue]{hyperref}

\usepackage[numbers,sort&compress]{natbib} 

\usepackage{bbm}
\usepackage[most]{tcolorbox}
\usepackage{fvextra}
\usepackage{xparse}
\usepackage{etoolbox}

\newcommand{\safeincludegraphics}[2][]{%
  \IfFileExists{#2}{%
    \includegraphics[#1]{#2}%
  }{%
    \fbox{\parbox{\linewidth}{\centering \textit{Missing figure:} \texttt{#2}}}%
  }%
}

\newtcolorbox{llmpromptbox}[2][]{%
  enhanced,
  breakable,
  colback=gray!5,
  colframe=gray!60,
  fontupper=\ttfamily\small,
  sharp corners,
  boxrule=0.4pt,
  title={#2},
  before skip=10pt,
  after skip=10pt,
  left=6pt,right=6pt,top=6pt,bottom=6pt,
  #1
}

\NewDocumentCommand{\prompt}{m o o}{%
  \edef\role{\IfValueTF{#2}{#2}{user}}%
  \edef\model{\IfValueTF{#3}{#3}{gpt-4}}%
  \IfFileExists{#1}{%
    \begin{llmpromptbox}{Role: \role, Model: \model}%
      \VerbatimInput[fontsize=\small, breaklines=true, obeytabs=true, tabsize=2]{#1}%
    \end{llmpromptbox}%
  }{%
    \begin{llmpromptbox}{Role: \role, Model: \model}%
      \textit{File not found:} \texttt{#1}. Place it next to this .tex or adjust the path.%
    \end{llmpromptbox}%
  }%
}

\usepackage{url} 

\newcommand{\TryFile}[3]{%
  \IfFileExists{#1}{#2}{%
    \IfFileExists{./#1}{#2}{%
      \IfFileExists{../#1}{#2}{%
        \IfFileExists{../../#1}{#2}{#3}%
      }%
    }%
  }%
}

\NewDocumentCommand{\modelOutput}{m o o}{%
  \par\medskip\noindent
  \begingroup
  \edef\initialState{\IfValueTF{#2}{#2}{user}}%
  \TryFile{#1}{%
    \begin{llmpromptbox}{Base hypothesis: \initialState}%
      \TryFile{#1}{%
        \VerbatimInput[fontsize=\small, breaklines=true, obeytabs=true, tabsize=2]{#1}%
      }{%
        \textit{Unexpected read failure for: }\path{#1}%
      }%
    \end{llmpromptbox}%
  }{%
    \begin{llmpromptbox}{Base hypothesis: \initialState}%
      \textit{Missing file: }\path{#1}%
    \end{llmpromptbox}%
  }%
  \endgroup
  \par\medskip
}
\title{\bfseries Tiny Moves: Game-based Hypothesis Refinement}

\author{
Agnieszka Dobrowolska\thanks{These authors contributed equally.},
Rogier Hintzen\footnotemark[1],
Martin Balla\footnotemark[1],
Karl Gemayel,\\
Sabine Reichert,
Thomas Charman,
Jen Ning Lim,
Lindsay Edwards,
Anna Gogleva\thanks{Corresponding author: \texttt{anna.gogleva@relationrx.com}}\\[0.5em]
Relation Therapeutics, London, UK
}

\date{} 
\begin{document}
\maketitle

\begin{abstract}
Most machine learning approaches to scientific discovery frame hypotheses as end-to-end predictions, obscuring the incremental structure of scientific reasoning. We propose The Hypothesis Game, a symbolic formalism for hypothesis refinement in which LLM agents operate on a shared hypothesis state using a fixed grammar of reasoning moves. The framework is motivated by the observation that scientific progress often proceeds through small, localized revisions, grounded in domain context, rather than extensive rewrites. We instantiate a minimal game with LLM agents and evaluate it on pathway-level mechanistic refinement tasks. In the primary setting of corruption recovery, where hypotheses contain controlled errors, the game-based approach consistently removes more errors and achieves higher precision than strong prompting baselines, while preserving valid structure through incremental edits. In a secondary reconstruction setting from partial cues, it performs comparably to the strongest baseline, indicating that explicit move-based refinement remains competitive even when ground-truth recovery is difficult. These findings support game-based reasoning as a principled route to more controllable, interpretable, and transferable hypothesis refinement systems for scientific discovery.
\end{abstract}

\section{Introduction}
Scientific discovery is rarely a single leap from data to conclusion. In empirical fields such as biology, the discovery process unfolds iteratively and non-linearly, often starting from partial hypotheses based on incomplete or noisy evidence. As new results emerge, researchers expand, revise, and combine hypotheses, allowing them to evolve over time. The emerging hypothesis undergoes multiple rounds of pruning, testing, and iterative refinement to converge on a coherent causal foundation \citep{Alkan2025survey}. In practice, this process frequently involves updating and reconciling partially correct hypotheses, rather than generating them entirely \textit{de novo}.

Recent work in AI for science has shown increasing interest in agentic approaches, where Large Language Models (LLMs) or multi-agent systems are assigned specialized roles, such as a literature reviewer, clinical trial designer, or experiment planner, to support parts of the scientific workflow \citep{Gridach2025-agentic, Zheng2025-autonomy}. Systems such as the "AI Co-Scientist" \citep{Gottweis2025-coscientist} and "Robin" \citep{Ghareeb2025-robin}, as well as lab-in-the-loop multi-agent frameworks \citep{Swanson2024-virtuallab} and domain-focused agent systems for biomedical discovery \citep{gao2024agents}, demonstrate how role-specific capabilities and tools can be orchestrated to address scientific problems end-to-end. While effective at targeting domain knowledge and tools, these systems typically emphasize task completion, with the internal process of scientific reasoning left largely implicit.

A central limitation of such approaches emerges when hypotheses must be revised rather than generated. Without explicit representations of intermediate hypothesis states or allowed transformations, it becomes difficult to localize errors, assess which parts of a hypothesis are well supported, or make targeted corrections. As a result, even minor inconsistencies or uncertainties are often addressed through broad, undifferentiated rewrites, reducing interpretability and control over the refinement process \citep{Mondorf2024-beyondaccuracy, Madaan2023-selfrefine}. This lack of structured refinement hinders the systematic reuse of reasoning patterns and limits transfer across related scientific problems \citep{Liu2023-agentbench, Majumder2024-discoverybench}. 

In contrast, human scientific reasoning is compositional: hypotheses are built and revised gradually from smaller fragments, guided by a repertoire of common reasoning patterns, such as expansion, pruning, comparison, and critique \citep{Lawson2004-scientific-reasoning}. Motivated by this observation, we propose a symbolic, game-based formalism for hypothesis refinement, in which LLM agents operate over a shared hypothesis state using a fixed grammar of reasoning moves. Each move defines an allowed transformation of the hypothesis, enabling refinement and evolution through localized interpretable edits, rather than broad rewrites. This conceptual framing makes the reasoning process explicit, allowing the system to “think about thinking” rather than hard-wiring problem-specific behaviours.

In this paper, we introduce \textbf{The Hypothesis Game}, a symbolic, game-based framework for hypothesis refinement. Our contributions are threefold. \textbf{First}, we formalise hypothesis refinement as a compositional reasoning game with a reusable grammar of moves, enabling hypotheses to be updated through localized, interpretable transformations rather than large-scale rewrites. \textbf{Second}, we present an implementation with LLM agents operating over shared hypothesis states, producing explicit refinement trajectories that support transparent and controllable reasoning. \textbf{Third}, we provide an empirical evaluation on pathway-level mechanistic refinement tasks derived from Reactome. We show that in a corruption recovery setting, where hypotheses contain controlled errors, the proposed framework consistently removes more errors and achieves higher precision than strong prompting baselines, while preserving valid hypothesis structure. In a complementary reconstruction-from-partial-cues setting, performance is comparable to the strongest baseline, highlighting both the strengths and limits of incremental refinement under partial information. Together, these results highlight the potential of game-based reasoning formalisms to support more granular, interpretable, and transferable scientific discovery.

\section{Framework}
The Hypothesis Game formalizes hypothesis refinement as the iterative transformation of a shared state through structured reasoning moves. This section defines how hypotheses are represented, how moves operate on them, and how modes and scoring functions may shape the dynamics of the game. 

Here we introduce a general operator-based formalism that captures a broad design space for hypothesis refinement. We intentionally instantiate only the minimal subset of this formalism required to evaluate the central research question: whether our proposed framework with a small, reusable reasoning grammar provides measurable benefits in biological refinement tasks. Other components, such as explicit scoring, policy-based or learned controllers, and richer hypothesis representations are optional extensions of the basic game. Our experiments are designed to isolate and test the general reasoning framework, while the broader formalism outlines how more sophisticated controllers and utilities can be incorporated in future work.

\begin{figure}[htbp]
  \centering
  \includegraphics[width=1\linewidth]{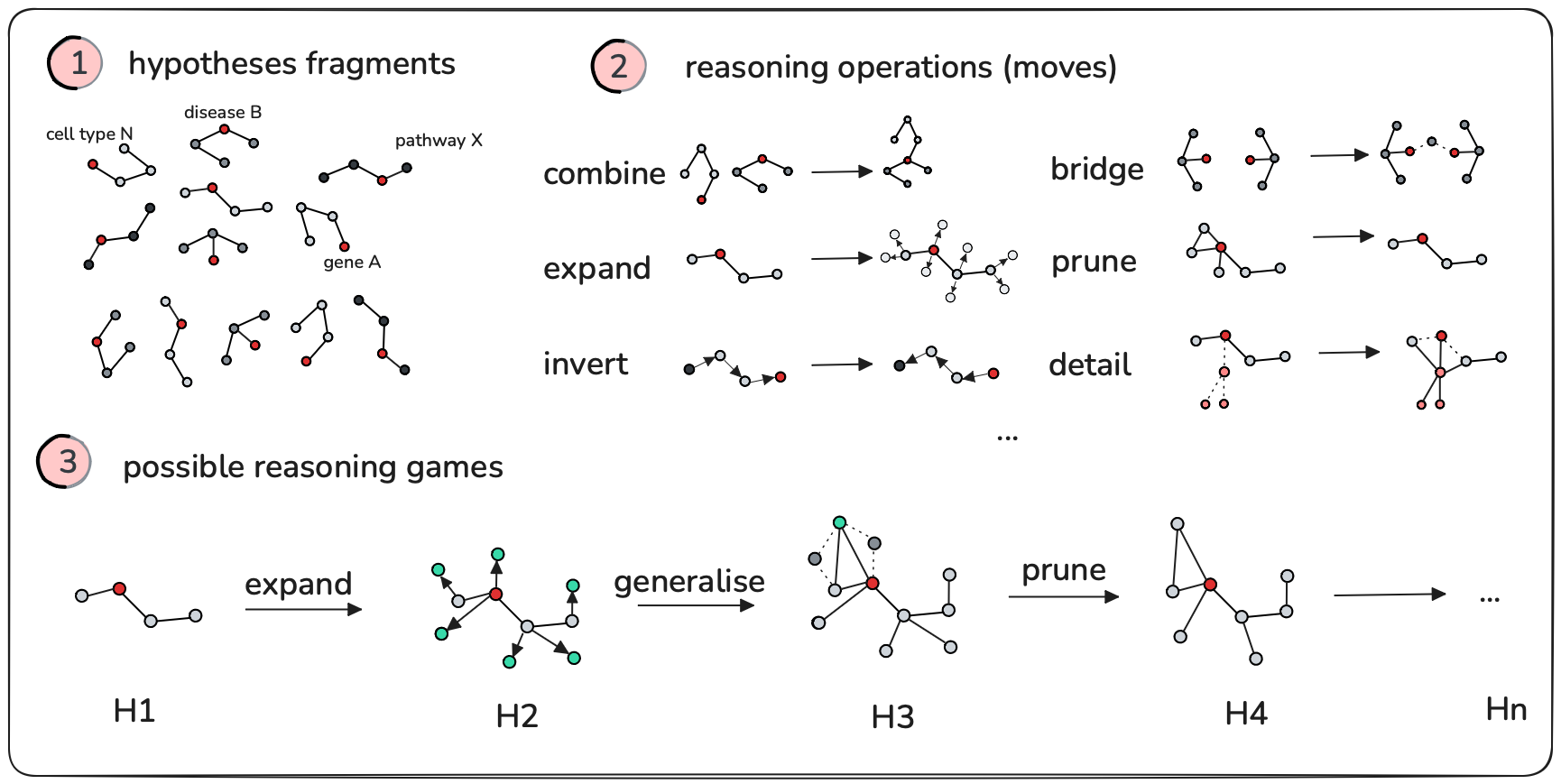}
  \caption{A conceptual framework for reasoning games. The objective of the game is to evolve a hypothesis fragment through a sequence of reasoning moves, with progress assessed through properties such as novelty, coherence, and traceability. \textit{*Graph structures shown for conceptual illustration only; actual implementation uses structured text fragments with equivalent reasoning operations.}
} 
  \label{fig:framework}
\end{figure}

\subsection{Hypothesis Representation}  
A hypothesis is represented as a set of fragments:  
\[
H_t = \{h_1, h_2, \ldots, h_n\},
\]  
where each fragment $h_i$ may be a text claim, a structured triple (subject–relation–object), or optionally mapped to a graph $G=(V,E)$ of entities and relations. In our experiments, we primarily use structured text.  

\subsection{Reasoning Grammar (Moves)}  

Let $\mathcal{O} = \{o_1, o_2, \dots, o_m\}$ denote a fixed set of reasoning operations.  
Formally, let $\mathcal{H}$ be the space of all possible hypotheses and $\mathcal{C}$ the space of contexts (e.g., cell type, disease, etc).  
Each operation is a function
\[
o_j : \mathcal{H} \times \mathcal{C} \mapsto \mathcal{H}, 
\quad (H_t, C) \mapsto H_{t+1},
\]
where $H_t \in \mathcal{H}$ is the current hypothesis, $C \in \mathcal{C}$ is an optional context (e.g., biological priors), and $H_{t+1} \in \mathcal{H}$ is the updated hypothesis state.

In our implementation, we restrict the set of moves to four core operations: \texttt{prune}, \texttt{expand}, \texttt{retrieve}, and \texttt{debate} (see Table \ref{tab:moves_summary}). Moves may be atomic (e.g. \texttt{prune}, \texttt{expand}) or composite (e.g. \texttt{retrieve\_expand}), consisting of a fixed sequence of atomic operations. More granular move types can be introduced as needed, typically informed by the structure of the underlying hypothesis representation. An example of a complete reasoning grammar based on graph representation of hypothesis fragments is shown in Fig.~\ref{fig:framework}.

Moves can be applied repeatedly and composed arbitrarily.  
We can define a maximum number of reasoning operations per round (move budget) as a fixed constant $k_{\text{max}}$. A round can be defined locally as one update step from $H_t$ to $H_{t+1}$, and globally, a sequence of rounds constitutes a complete game. 
\[
H_{t+1} = \left( \prod_{i=1}^{k} o_{j_i} \right)(H_t, C), \quad k \le k_{\max}.
\]  

At each round, a controller selects and applies up to $k_{\text{max}}$ moves to evolve the hypothesis. The controller can be realized in different ways (e.g., an LLM, finite state machine, or RL agent), depending on the desired game design.

\subsection{Game Modes}  
In open-ended discovery, the precise outcome is often unknown, but the overall style of reasoning can still be guided. We capture this through a \emph{mode} $M$, which specifies how moves are selected. One way to formalise this idea is through a probability distribution over moves,  
\[
\pi_M(o_i \mid H_t) = P(\text{apply } o_i \mid M),
\]  
where, for example, a \textit{discovery} mode favors generative moves such as \texttt{expand}, while a \textit{validation} mode favors critical moves such as \texttt{prune} or \texttt{debate}. More generally, modes can also be realized by restricting the available moves $\mathcal{O}$, enforcing deterministic rules, or combining weighting and constraints set by the overall objective of a game.  

In our experiments, modes are approximated through natural language instructions to the controller, but the reasoning grammar provides a principled way to configure high-level exploration or validation goals in more open-ended settings.

\subsection{Scoring} 
While modes can guide reasoning styles at a high level, scoring functions may offer a way to make the game more controllable. Quantifying metrics during refinement provides a way to shape the trajectory of the game. Formally, we can define a vector of metrics,  
\[
S(H_t) = \left( D_{\text{known}}(H_t),\ \Delta_{\text{div}}(H_t),\ L_{\text{connect}}(H_t),\ T_{\text{frag}}(H_t) \right),
\]  
where the components capture distance from known hypotheses ($D_{\text{known}}$), diversity of current hypothesis ($\Delta_{\text{div}}$), local connectivity ($L_{\text{connect}}$), and traceability to prior knowledge ($T_{\text{frag}}$).  
These can be aggregated into a scalar utility,  
\[
U(H_t) = \beta^\top S(H_t),
\]  
with weights $\beta$ reflecting mode-specific priorities (e.g., traceability in \textit{validation}, diversity in \textit{discovery}). In practice, robust scoring for biological hypotheses likely requires a \emph{hybrid} setup that combines computational metrics with experimental signals, even if sparse. For example, hypothesis fragments involving molecular interactions could be evaluated using targeted binding assays or perturbation readouts, providing grounded feedback that complements algorithmic metrics.

In this work, we do not use explicit scoring to drive the controller; modes are implemented through natural-language instructions. The scoring framework presented here is therefore conceptual, illustrating how computationally and experimentally informed metrics could be integrated into more autonomous implementations in the future.

\subsection{Game variants}  
The outlined game formalism allows us to define game variants that operate on different granularity levels. \textbf{Simple Hypothesis Refinement} treats the whole hypothesis as a single state (\autoref{alg:simple}). In each round, a mode-conditioned controller selects a move from the shared grammar and updates the entire state, stopping when task goals are met. 

\begin{algorithm}[H]
\caption{Simple Hypothesis Refinement (single round)}
\label{alg:simple}
\begin{algorithmic}
\Require initial hypothesis state $H_0$, reasoning moves $\mathcal{O}$, mode $M$, move budget $k_{\max}$, context $C$, termination criteria
\State $t \gets 0$
\While{\texttt{not Terminate}($H_t$)}
    \State \textbf{Game Master:} provide current state $H_t$ and mode $M$ to controller
    \State \textbf{Controller:} select sequence of moves $(o_{j_1},\dots,o_{j_k})$ with $k \le k_{\max}$ according to $\pi_M$
    \For{each $o_{j}$ in selected moves}
        \State $H_t \gets o_j(H_t, C)$ \Comment{apply reasoning move with optional context $C$}
    \EndFor
    \State $t \gets t+1$
\EndWhile
\State \Return final hypothesis $H_t$
\end{algorithmic}
\end{algorithm}

Noting that large changes are rarely necessary to refine a hypothesis, we can build on the simple variant by enabling granular edits during the hypothesis’ evolution. \textbf{Localized Hypothesis Refinement} keeps the same controller and move set but operates on fragments (structured text or subgraphs), selecting regions to edit and enforcing global consistency so untouched parts remain unchanged (\autoref{alg:localized}). This game type strongly depends on the underlying hypothesis representation structure. 

\begin{algorithm}[H]
\caption{Localized Hypothesis Refinement (single round)}
\label{alg:localized}
\begin{algorithmic}
\Require Hypothesis state $H_t=\{h_1,\dots,h_n\}$ (structured text or graph), moves $\mathcal{O}$, mode $M$, move budget $k_{\max}$, context $C$, selector $\sigma$
\State \textbf{Selector} $\sigma$: propose a set of candidate regions $\mathcal{R}=\{R_1,\dots,R_m\}$ where each $R_i \subseteq$ nodes/tuples of $H_t$
\State \textbf{Controller} (mode $M$): choose up to $k \le k_{\max}$ pairs $\{(o_j,R_j)\}_{j=1}^k$ with $o_j \in \mathcal{O}$
\For{each $(o_j,R_j)$}
  \State $H_t \gets \texttt{ApplyLocal}(H_t, o_j, R_j, C)$ \Comment{local rewrite on $R_j$ only}
  \State $H_t \gets \texttt{EnforceConsistency}(H_t, R_j)$ \Comment{maintain schema/typing/acyclicity/etc.}
\EndFor
\State \Return $H_t$
\end{algorithmic}
\end{algorithm} 

Together, these variants illustrate that the formalism supports both high-level, whole-state reasoning and fine-grained, region-focused reasoning under a shared utility definition and mode settings. The simple variant is recovered when the selected region spans the full state. This design mirrors the varying levels of complexity observed in biological systems.

\section{Implementation} 
To test the proposed framework, we implement a minimal version of the game as a system of specialized agents, where the reasoning process is determined by a central LLM controller, \textbf{Game Master}. The Game Master guides the reasoning process by iteratively analyzing the hypothesis state and selecting moves based on the analysis. Move selection consists of a clear request (e.g. \textit{"remove component A from the hypothesis"}) and which agent(s) should execute it. Table~\ref{tab:moves_summary} summarizes the moves, their components, and corresponding responsibilities. In the implementation, the abstract \texttt{retrieve\_expand} move is instantiated in two variants, depending on the source of evidence. 

\begin{table}[ht]
\centering

\caption{Key elements of The Hypothesis Game. 
Full prompts are provided in the Supplementary Methods
(see Section~\ref{sec:suppmethods-moves-prompts}).}

\begin{tabular}{M{2.8cm} L{3cm} L{7.5cm}}
\toprule
\textbf{Move} & \textbf{Components} & \textbf{Description} \\
\midrule
\multirow{2}{*}{\parbox[c]{\linewidth}{\raggedright\textbf{Game Master (LLM controller)}}}
  & Diagnose & Evaluate hypothesis and recommend next actions. \\
  & Move selection  & Choose next move based on recommendations. \\
\midrule
\textbf{Prune}
  & Prune & Remove component(s) from hypothesis. \\
\midrule
\multirow{2}{*}{\parbox[c]{\linewidth}{\raggedright\textbf{Expand with corpus}}}
  & Retrieve evidence & Search external corpora for evidence. \\
  & Expand & Integrate retrieved information into the hypothesis. \\
\midrule
\multirow{2}{*}{\parbox[c]{\linewidth}{\raggedright\textbf{Expand with LLM introspection}}}
  & Retrieve evidence & Gather information using LLM prior knowledge. \\
  & Expand & Integrate retrieved information into the hypothesis. \\
\midrule
\multirow{3}{*}{\parbox[c]{\linewidth}{\raggedright\textbf{Debate}}}
  & Setup & Frame the debate around the requested topic. \\
  & Debate topic & Multiple agents argue from distinct positions. \\
  & Conclude & Analyse the debate and propose a final conclusion. \\
\bottomrule
\end{tabular}
\label{tab:moves_summary}
\end{table}

\textbf{Modes:} In our minimal prototype, modes are realized by injecting mode descriptions into the initial prompt to the Game Master (controller). This prompt influences the choice of reasoning operations without an explicit probabilistic policy module. While simplified, this approach provides a controllable approximation of $\pi_M$ and allows us to explore the impact of different modes.

\textbf{Game control:} Game goals and stopping conditions are specified to the Game Master (controller) through the initial prompt, and the Game Master's \textit{Diagnose} component decides when the hypothesis satisfies the requirements. Although this approach lacks explicit metric-based control, it provides a flexible mechanism for steering the game. The scoring function described above is presented as part of the general formalism, illustrating how automated, quantitative evaluation could be incorporated in future implementations.

\section{Experiment set-up}
Reasoning benchmarks in mathematics and common sense (GSM8K \citep{cobbe2021gsm8k}, MATH \citep{hendrycks2021math}, BIG-Bench \citep{srivastava2022bigbench}) do not directly translate to biological hypothesis construction and refinement, where researchers must build complex hypotheses step by step from incomplete, noisy, sometimes contradictory evidence rather than retrieving facts. Without established ways to evaluate reasoning quality, benchmarks should challenge systems to tolerate noise, recover missing links, and extend hypotheses in controlled ways. Emerging biological benchmarks such as BioMaze \citep{zhao2025-biomaze} move in this direction with graph-based pathway QA and high-level LLM-as-judge evaluations, but still differ from the longer-horizon, statement-level refinement studied here.

To fill this gap, we introduce two evaluation tasks designed as first benchmarks for hypothesis refinement. These tasks mirror realistic challenges in biological discovery: (1) corruption recovery and (2) hypothesis reconstruction from partial cues  (Table~\ref{tab:evaluation-tasks}). 

\begin{table}[H]
  \caption{Evaluation tasks overview}
  \label{tab:evaluation-tasks}
  \centering
  \begin{tabular}{p{2cm} p{3.7cm} p{3.2cm} p{2.5cm}}
    \toprule
    \multicolumn{1}{c}{\textbf{Task}} &
    \multicolumn{1}{c}{\textbf{Purpose}} &
    \multicolumn{1}{c}{\textbf{Validates}} &
    \multicolumn{1}{c}{\textbf{Metrics}} \\
    \midrule
    Corruption Recovery &
    Can the system correct noisy or misleading hypotheses? &
    Robustness to noise; Logical refinement &
    Error removal rate, precision, recall, F1 \\
    \addlinespace[2pt]
    Reconstruction from Partial Cues &
    Can the system rebuild known mechanisms from partial cues? &
    Incremental reasoning; Traceability &
    Precision, recall, F1 \\
    \bottomrule
  \end{tabular}
\end{table}

\begin{figure}[htbp]
  \centering
  \includegraphics[width=1\linewidth]{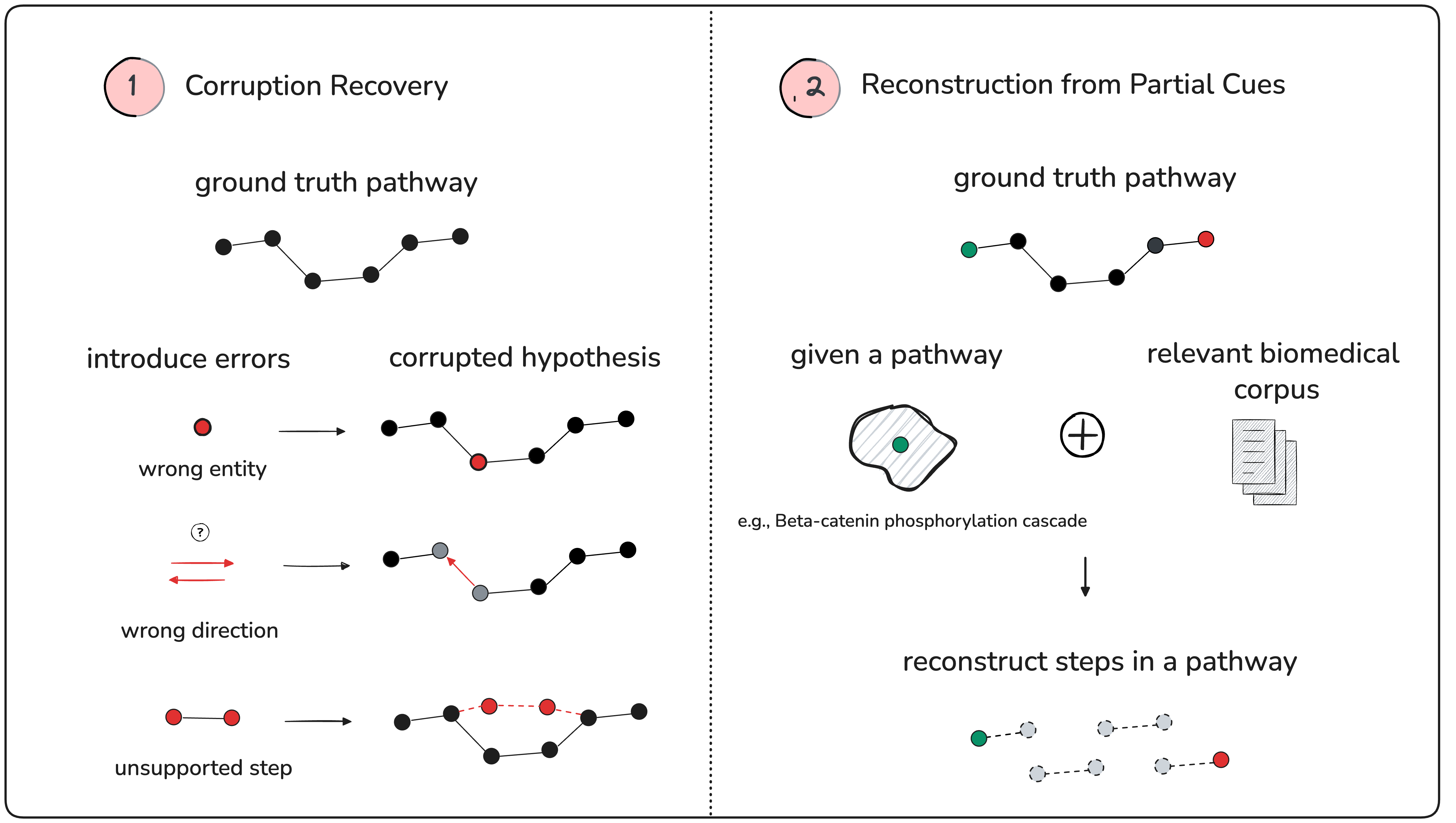}
  \caption{Conceptual illustration of the two evaluation tasks. \textit{Left:} corruption recovery, where controlled errors are introduced into a valid pathway and the system must detect and repair them while preserving correct structure. \textit{Right:} reconstruction from partial cues, where a system recovers pathway steps starting from sparse contextual input and external biomedical evidence.
} 
  \label{fig:tasks}
\end{figure}

Figure~\ref{fig:tasks} provides a conceptual illustration of the two evaluation tasks, while Table~\ref{tab:evaluation-tasks} summarizes their evaluation goals and metrics.

\subsection{Task Setup}
We instantiate evaluation tasks using curated subsets of human pathways from Reactome \citep{Jassal2020-reactome}. Each pathway consists of biochemical reactions, available in both graph and text representations (see \ref{sec:suppmethods-reconstruction-dataset}). In the text representation, pathways are expressed as sets of statements describing biochemical reactions; for example, \textit{ATP phosphorylates glucose to form glucose-6-phosphate.}

We sampled pathways stratified by the number of biochemical reactions, to capture the diversity and complexity of the complete dataset. For reconstruction and corruption tasks, we sampled 100 and 20 pathways, respectively. The rationale was to create datasets large enough to capture key reasoning patterns across multiple approaches, while remaining feasible for large-scale experimentation. In total, we ran 820 experiments for reconstruction and 2880 experiments for corruption. Corruption tasks test multiple controlled variants per pathway, which resulted in a larger number of individual experiments. 

\paragraph{Common Experimental Principles}
Across all tasks, hypotheses are represented as text fragments. The Hypothesis Game is restricted to four available moves: \texttt{prune}, \texttt{expand} (with corpus or with LLM introspection), and \texttt{debate} (See Table~\ref{tab:moves_summary}). Move selection and termination are dynamically governed by the Game Master, adapting to task-specific goals.

We compared our approach against three reasoning baselines: Zero-Shot prompting, Chain-of-Thought, and ReAct. Zero-Shot directly generates answers without intermediate reasoning steps \citep{Brown2020-gpt3}. Chain-of-Thought elicits step-by-step reasoning through intermediate natural language explanations \citep{Wei2022-cot}. ReAct interleaves reasoning traces with access tools to improve decision making \citep{Yao2022-react}. We compared these baselines against our Hypothesis Game under different move configurations and a fixed move budget. All models received the same input prompt (see Supplementary A Sec. \ref{sec:suppmethods-moves-prompts}), which instructs the system to either reconstruct a pathway or recover a corrupted pathway. All curated datasets are available on Hugging Face\footnote{\url{https://huggingface.co/datasets/TuringRRX/TinyMoves}}.

\textbf{Task 1 – Corruption Recovery:}
The corruption task assesses the ability to detect and repair errors while preserving the structure of a valid pathway. Starting from 20 human pathways, we introduced three types of corruptions (errors) (See Supplementary Table~\ref{tab:corruption-dataset-design}): 

\begin{itemize}\setlength\itemsep{0pt}
    \item wrong entity – replacing a correct entity with an incorrect one; 
    \item wrong relationship – altering the relation between entities; 
    \item irrelevant statement – inserting a non-relevant statement into the pathway. 
\end{itemize}

We further varied the level of challenge along two axes: 1) \textbf{difficulty:} \textit{easy} (trivial errors) and \textit{hard} (subtle changes, requiring a deeper biological understanding);  2) \textbf{error rate:}  10-40\% of pathway length (measured as a number of steps/reactions) to capture differences in pathway size and complexity. 
All errors were generated by an LLM and iteratively refined, with two domain experts reviewing and manually correcting outputs to produce the curated corruption set.

\textbf{Evaluation} combined two measures. First, an LLM judge was presented with the original statement, the corrupted version, and the model’s output, and determined whether the error persisted. Second, entity mapping, as in reconstruction, quantified biological fidelity by measuring precision and recall of annotated entities against the ground truth.

\noindent
\textbf{Task 2 – Reconstruction from Partial Cues:}
The reconstruction task evaluates whether a system can reconstruct complex hypotheses from partial cues by performing incremental reasoning. Starting from a minimal cue, the system must recover the biochemical reactions (steps) of a biological pathway, modeling the onerous curation process domain experts go through to construct the Reactome database. To reduce the risk of models exploiting memorized knowledge of well-known pathways, we rephrased pathway names while preserving their semantic content and level of granularity. A domain expert inspected and corrected the paraphrased titles to ensure semantic fidelity (available on \href{https://huggingface.co/datasets/TuringRRX/TinyMoves/blob/main/task_1_reconstruction/pathway_name_mapping.tsv}{Hugging Face.)}
For agents with tool access (our approach and ReAct), we additionally provided a corpus of open-access biomedical articles, consisting mainly of abstracts cited in the Reactome pathway descriptions.

\textbf{Evaluation} relied on two complementary notions of correctness. At the pathway level, we annotated entities (genes, protein complexes/families, and chemicals) in both original and generated pathways using Gilda~\citep{gyori2022gilda}; precision and recall over these entity sets provided a quantitative measure of biological fidelity. At the reaction level we refer to the LLM-as-judge metric as `Detailed Recall', it evaluates whether the generated pathways reproduced the intended biochemical reactions, assessing four attributes: input entities, output entities, reaction directionality, and type of biological interaction (Supplementary A Sec. \ref{sec:suppmethods-moves-prompts}). To assess the reliability of the LLM-as-judge, we conducted a post-hoc calibration study in which two senior domain experts independently scored a stratified sample of model outputs for both tasks (Supplementary A Sec. \ref{supp-llm-as-judge}).

\section{Results}
We evaluated \emph{The Hypothesis Game} on two pathway-level reasoning tasks described above: recovery of corrupted hypotheses and reconstruction from partial cues. In both settings, we compare the \emph{Hypothesis Game} configuration (four move types with access to the corpus) against strong prompting baselines (Zero-Shot, Chain-of-Thought, ReAct). Our analysis focuses on a minimal instantiation of the game designed to isolate the effects of incremental, move-based refinement. The underlying formalism naturally extends to move sets and reasoning modes.

\begin{figure}[h]
  \centering
  \includegraphics[width=1.0\textwidth]{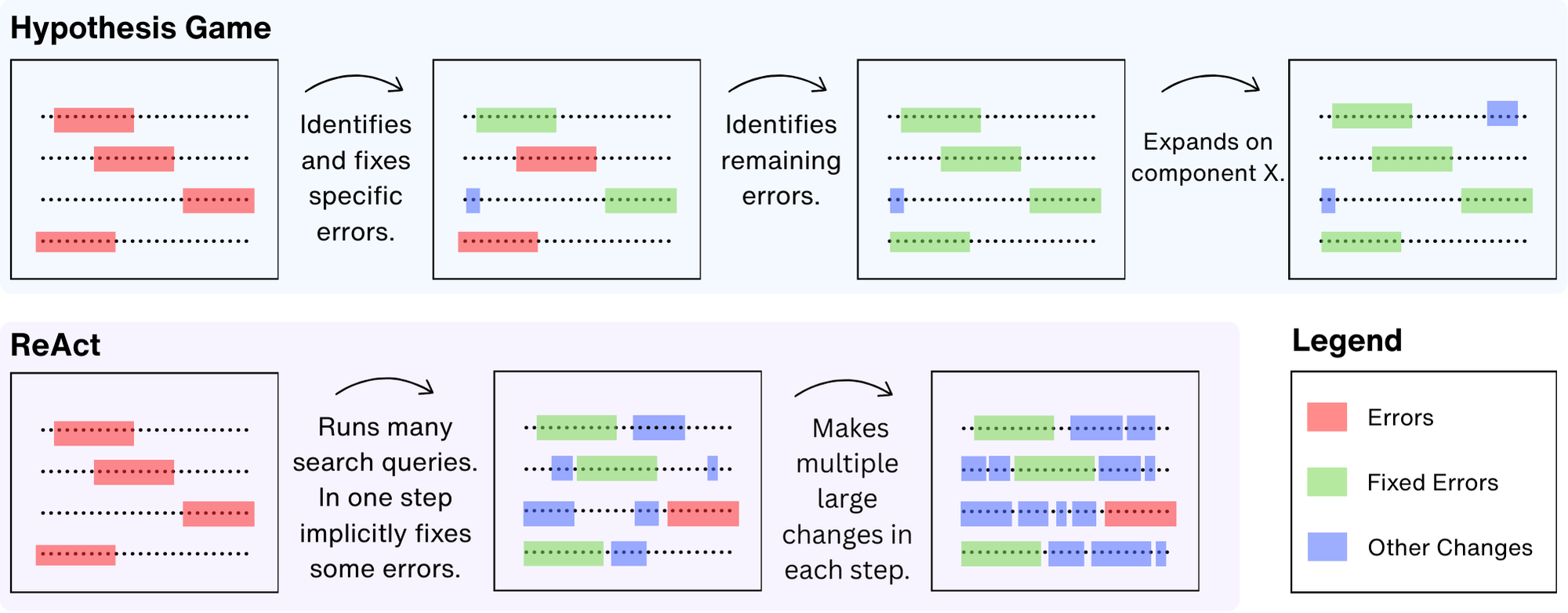}
  \caption{Representative example run of \textit{Hypothesis Game} and ReAct on the corruption task, illustrating incremental vs large single-step edits. \textit{*Other changes} are quantified as (1) the number of biological entity additions/removals and (2) word-level normalised Levenshtein distance to the reference pathway. See Fig.~\ref{fig:hypothesis_drift} for details.}
  \label{fig:eye_candy}
\end{figure}

\paragraph{Qualitative observations.} In the corruption recovery task, \emph{The Hypothesis Game} exhibits a distinctive pattern of incremental and traceable edits. As illustrated in Figure ~\ref{fig:eye_candy}, the game progressively identifies and corrects individual errors while making only minor additional changes to the input hypothesis. In contrast, ReAct tends to modify the pathway through fewer but substantially larger updates, often overwriting multiple components at once and introducing much larger deviations from the original hypothesis structure. Quantitative summaries of such changes are shown in Fig.~\ref{fig:hypothesis_drift}. 

We observed a similar pattern, though less pronounced, in the reconstruction task. There, \emph{The Hypothesis Game} again favors smaller, localized updates, while the prompting baselines often overwrite large portions of the intermediate hypothesis in a single step (see Supplementary B Sec.~\ref{sec:suppmethods-reconstruction-game-example} for a complete example). Together, these observations highlight the role of controlled step-by-step refinement in maintaining hypothesis coherence across both tasks. 

\begin{figure}[h]
   \centering
   \includegraphics[width=\linewidth]{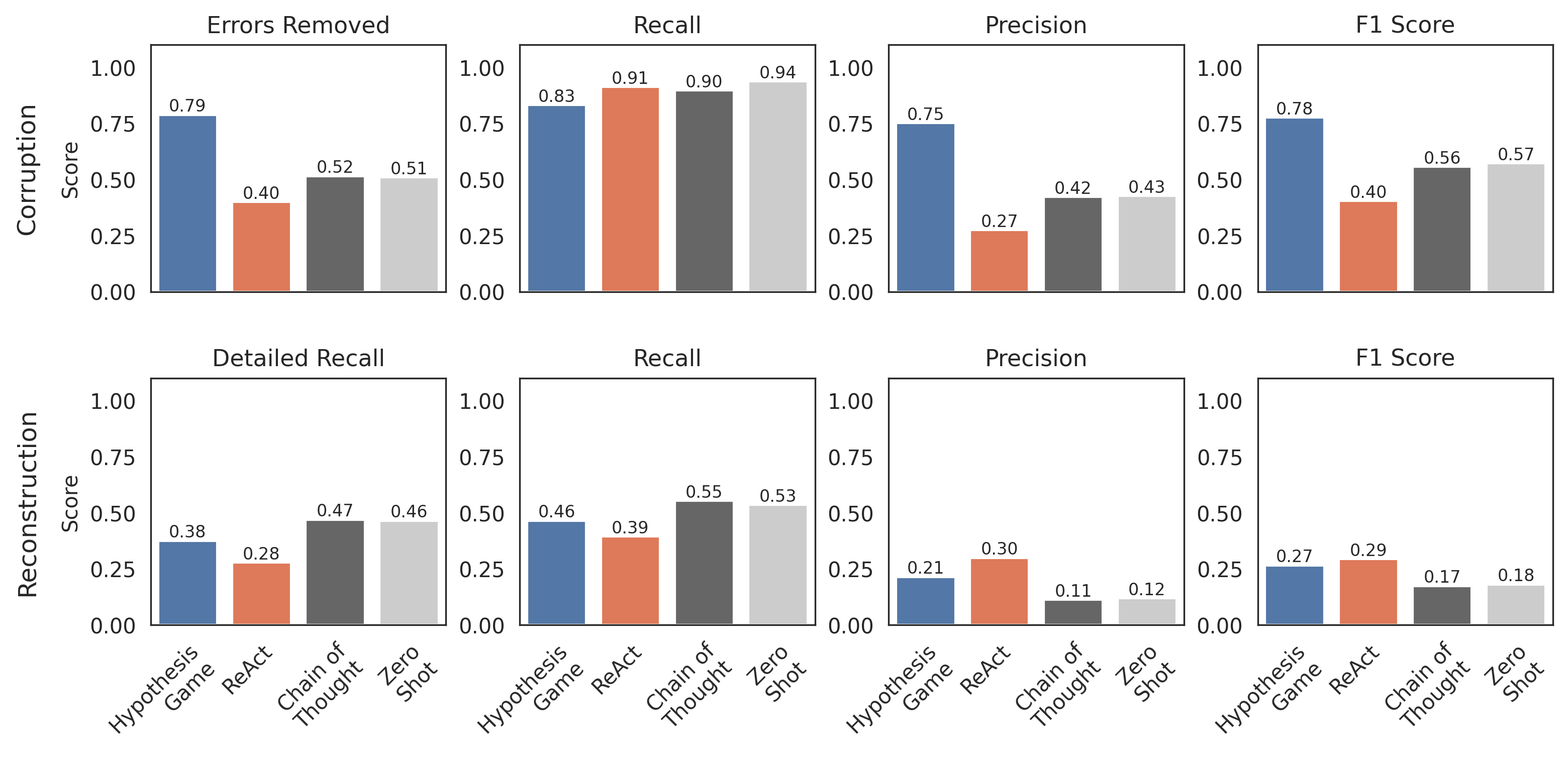}
    \caption{Comparison of \textit{Hypothesis Game} vs. prompting baselines on two pathway-level tasks. Bars show averages over the evaluation sets described in the text. The error bars show 95\% confidence intervals. 
    Top row: \textbf{Corruption}; \textit{Hypothesis Game} balances error removal and retention of valid content, achieving the highest precision, F1 and error removal rate (for all scores Friedman test $p<0.0001$, post-hoc Wilcoxon test with Bonferroni correction $p<0.0005$). Bottom row: \textbf{Reconstruction}; All methods struggled with faithfully reconstructing the pathways. \textit{ReAct} and \textit{Hypothesis Game} had a statistically non-significant difference in F1 score, but \textit{Hypothesis Game} performed significantly better in Detailed Recall of pathways (Friedman test, $\chi^2(3)=84.3, p<0.0001$, post-hoc Wilcoxon test with Bonferroni correction $p < 0.001$).  }
    \label{fig:combined_dummy}
\end{figure}

\paragraph{Corruption task.}
In the corruption recovery task (error rates 10–40\%), \emph{The Hypothesis Game} achieves the strongest overall performance across all evaluated metrics. Fig.~\ref{fig:combined_dummy} summarises results aggregated across pathways, corruption types, and error rates. The \textbf{Errors Removed} panel shows that \emph{Hypothesis Game} consistently removes a larger fraction of injected errors than all prompting baselines.

The \textbf{Recall} and \textbf{Precision} panels further highlight a characteristic trade-off between methods. ReAct achieves high recall by aggressively modifying the hypothesis, but this comes at the expense of precision, as many valid components are overwritten or altered. Chain-of-Thought and Zero-Shot prompting, in contrast, tend to preserve existing content, but fail to reliably remove corrupted statements, resulting in lower error removal rates. \emph{The Hypothesis Game} balances these extremes, combining strong error removal with the highest precision and overall \textbf{F1 score}, selectively correcting corrupted components while preserving valid pathway structure.

\begin{figure}[h]
    \centering
    \includegraphics[width=\textwidth]{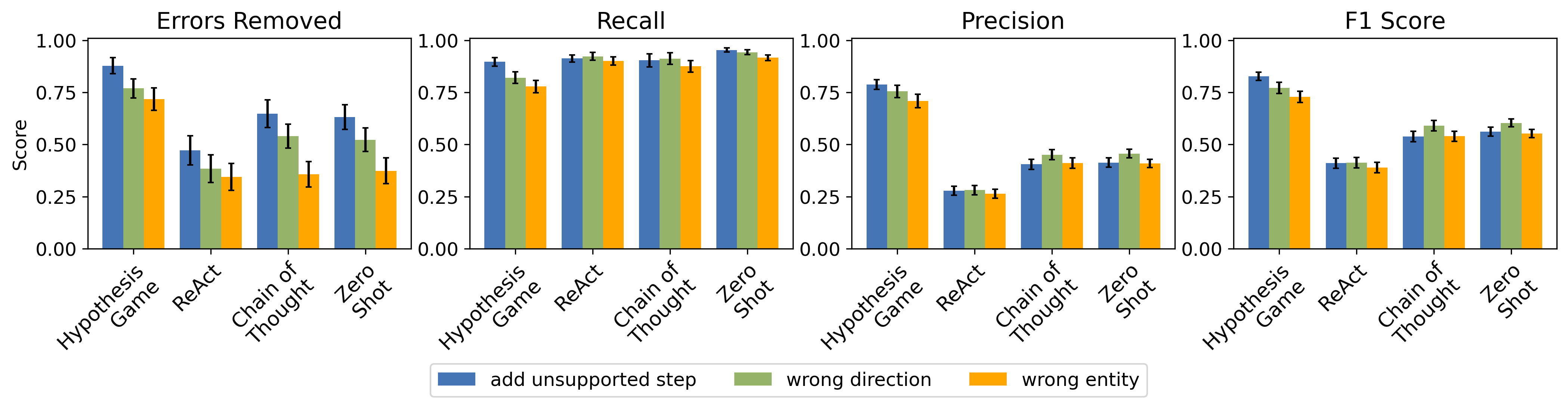}
    \caption{Aggregation of all results on the corruption task based on error type. Error bars show 95\% confidence intervals.}
    \label{fig:error_type}
\end{figure}

A breakdown by corruption type provides additional insight into this behavior (Fig.~\ref{fig:error_type}). \emph{Unsupported step} errors are most easily removed, as they introduce entire statements that are readily identified as irrelevant. \emph{Wrong-direction} corruptions are more challenging, since they preserve surface plausibility while inverting causal polarity. \emph{Wrong-entity} substitutions prove the most difficult: the resulting pathways remain fluent but introduce subtle inconsistencies in biochemical grounding, requiring deeper semantic discrimination. Across all corruption types, \emph{The Hypothesis Game} achieves the strongest performance, with particularly large gains on entity- and relationship-level errors. 

The complete results, stratified by corruption difficulty and fraction of injected errors, are provided in Supplementary B Sec.~\ref{sec:coruption-stratified-results}.

\paragraph{Reconstruction task.}
Reconstruction from partial cues represents a substantially less constrained setting than corruption recovery, requiring the system to infer missing pathway components from sparse contextual information. In this setting, \emph{The Hypothesis Game} performs comparably to the strongest baseline, ReAct, and outperforms Zero-Shot and Chain-of-Thought prompting (Fig.~\ref{fig:combined_dummy}).

While some Reactome pathways are relatively well known, allowing LLMs to recall key components, this advantage primarily manifests as higher recall for Chain-of-Thought and Zero-Shot prompting. However, these methods frequently introduce a large number of additional entities and reactions not present in the reference pathway, resulting in substantially lower precision (Fig.~\ref{fig:added-entities}). In contrast, \emph{The Hypothesis Game} maintains tighter control over hypothesis growth, favoring incremental expansion over broad completion.

Overall, ReAct achieves slightly higher F1 scores than \emph{The Hypothesis Game}, followed by Zero-Shot and Chain-of-Thought prompting. Nevertheless, low precision–recall values across all methods highlight the intrinsic difficulty of the reconstruction task. Beyond the inherent challenge of reconstructing pathways typically curated by domain experts, performance is further limited by incomplete information in partial cues, heterogeneity in pathway definitions, and the abstract-biased nature of the underlying biomedical corpus.

To better understand the contribution of individual reasoning operations in this setting, we conducted an ablation study over all subsets of the four core moves, as well as removing access to the external corpus (20 pathways; Table~\ref{tab:ablation-combined}). This analysis clarifies which move types are most critical for incremental reconstruction under sparse information, and further illustrates the limits of the minimal game configuration in open-ended settings.

\paragraph{Summary}
Our results highlight complementary strengths across the two tasks. In corruption recovery, the advantages of structured reasoning are evident: \emph{Hypothesis Game} achieved the highest overall performance, combining strong error removal with superior precision and F1 scores, while maintaining recall. In reconstruction from partial cues, all methods struggled to recover complete pathways, reflecting the inherent difficulty of this setting. Here, the \emph{Hypothesis Game} matched the strongest baseline (ReAct), while outperforming simpler prompting strategies in precision. Taken together, these findings suggest that the game-based framework, centered on small incremental reasoning steps ("tiny moves"), is particularly effective in settings that require targeted error correction and robustness to noisy inputs. 

\section{Conclusions and Future Work}

Our study shows that a structured, game-based approach to hypothesis refinement can match strong prompting baselines in reconstruction tasks and clearly outperform them in corruption recovery. In the latter setting, explicit reasoning moves enable targeted error correction while preserving valid pathway structure, yielding higher precision and overall repair quality.  Together, these results highlight both the promise and the limitations of current methods: while controlled corruption recovery benefits strongly from structured reasoning, open-ended reconstruction remains a challenging setting for all approaches. Although our experiments focus on settings with known ground truth, the proposed formalism is not inherently limited to consistency-bound refinement and can, in principle, support more exploratory forms of hypothesis evolution.

Future work will extend the framework along several directions. First, we will explore richer hypothesis representations, including structured and semi-structured text and graph formalism. Second, we plan to investigate metric-driven and learned controllers for move selection, building on the conceptual scoring framework introduced here. Third, we aim to broaden the evaluation suite to include open-ended hypothesis evolution scenarios.  
Taken together, these directions move from controlled settings with known ground truth toward more realistic discovery scenarios, enabling both consistency-driven refinement and more exploratory reasoning where robustness, novelty, and interpretability are critical.

\subsubsection*{Acknowledgments}
We thank our colleagues and collaborators for their support and constructive feedback during this work: Jake Taylor-King, Thomas Gaudelet, Alex Zhebrak, Anna Mastela, Eigenia Sergeev, Giannis Loukas, Shabbir Khan, Cristian Regep, Sarah-Jane Dunn. We also thank the Reactome team for making the pathway knowledge base openly available.

\clearpage
\appendix 

\section{Reconstruction}

\subsection{Dataset Creation}
\label{sec:suppmethods-reconstruction-dataset}

We filtered the Reactome database to ``leaf'' pathways—those that contained no other pathways nested within them—and stratified them into 10 bins based on the number of reactions per pathway. From these 10 bins we sampled 10 pathways for a dataset of 100 pathways.\newline

To assemble a relevant corpus for the reconstruction task we extracted the annotated \texttt{Publication Reference} from each of the sampled pathways in the Reactome database. For each pathway we then downloaded a corpus of articles based on the document identifiers. For the vast majority of articles we were only able to download an abstract due to their copyright license limiting their distribution (85\% abstract-only, 13\% full text, 2\% unavailable).\newline

\begin{figure}[h]
    \centering
    \safeincludegraphics[width=\textwidth]{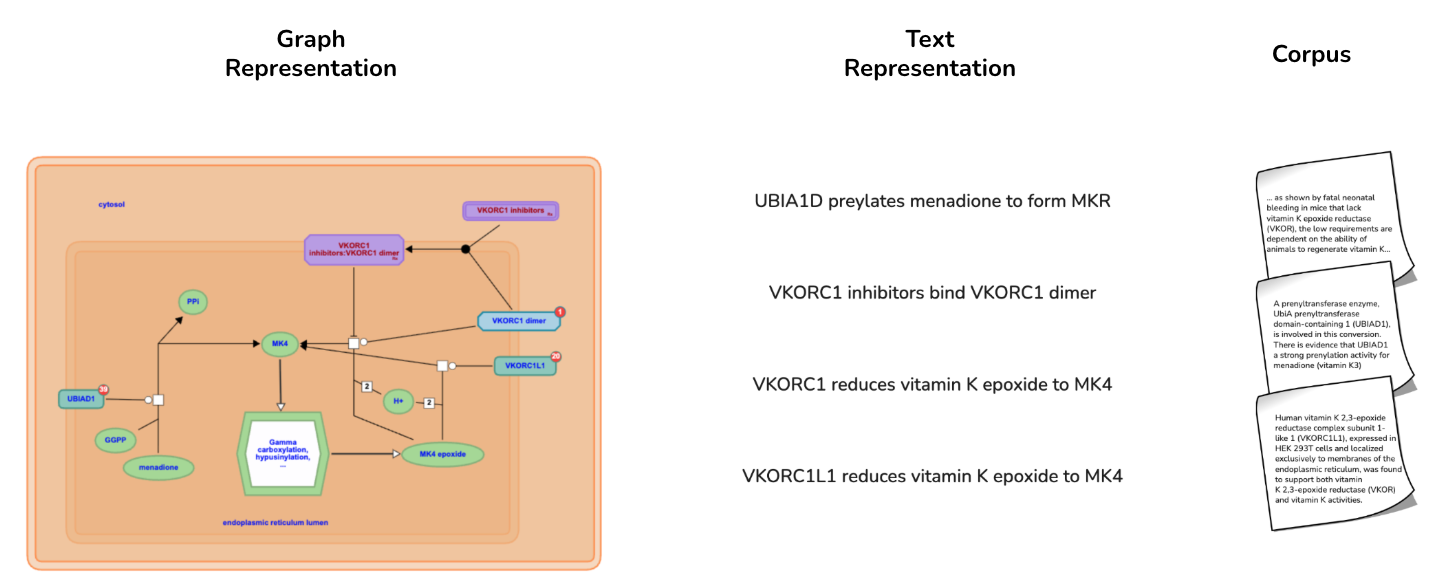}
    \caption{Example of a Reactome pathway (\protect\href{https://reactome.org/PathwayBrowser/\#/R-HSA-6806664\&SEL=R-HSA-6806647\&PATH=R-HSA-1430728,R-HSA-196854,R-HSA-6806667}{R-HSA-6806667}), displaying the full graph representation, the text representations of the biochemical reactions, and the associated corpus.}
    \label{fig:reactome_pathway}
\end{figure}

\newpage

\section{Corruption}
\subsection{Dataset Creation}
\label{sec:suppmethods-corruption-dataset}

We construct a controlled dataset of systematically corrupted pathways. The process has three stages:
\begin{enumerate}
\item \textbf{Corruption bank}. For each pathway in the reference set, and for each individual step, we pre-generate candidate corruptions across all error categories (wrong entity, wrong relation, unsupported step) and both difficulty levels (easy, hard). This ensures full coverage of possible perturbations. The specifications for creating the corruptions bank are shown in Table \ref{tab:corruption-dataset-design}. Candidate corruptions were
first generated by an LLM and then iteratively curated in collaboration with two domain
experts, who reviewed multiple rounds of generations and manually corrected remaining issues until
each example satisfied the intended error type and difficulty.

\item \textbf{Sampling policy}. A deterministic sampling script then assembles corrupted pathways by selecting (i) a target error category, (ii) a difficulty level, and (iii) a fraction of steps to corrupt. Importantly, only one corruption is allowed per step, guaranteeing that evaluation isolates the effect of single errors rather than compounded noise.

\item \textbf{Application}. Given these specifications, the corruption plan is applied to the pathway: original steps are replaced or augmented according to the corruption metadata, and both the corrupted pathway and detailed metadata (anchor indices, operation type, corrupted text) are saved. Random seeds make the process reproducible and allow controlled variation across runs. The corrupted pathways along with the metadata are available online at \url{https://huggingface.co/datasets/TuringRRX/TinyMoves}.
\end{enumerate}

This design yields a benchmark where the exact location, type, and difficulty of each corruption is known. By controlling error density and forbidding multiple corruptions per step, the dataset provides a clean experimental environment for measuring whether systems can remove or withstand specific classes of noise without conflating them.

\begin{table}[h]
  \caption{Corruption dataset design}
  \label{tab:corruption-dataset-design}
  \centering
  \small
  \renewcommand{\arraystretch}{1.15}
  \begin{tabularx}{\textwidth}{lXXX}
    \toprule
    & \textbf{Wrong entity} & \textbf{Wrong relation} & \textbf{Unsupported step} \\
    \midrule
    \textbf{Type} & Modify existing step & Modify existing step & Add a new step \\
    \textbf{Operation} & Replace (swap exactly one entity; verb unchanged) & Replace (keep entities; change verb or polarity) & Insert (add new statement) \\
    \textbf{Description} & Wrong entity (gene, protein, complex, isoform, state species) substituted into an otherwise valid step. & Entities unchanged, but relationship inverted (subject--object, activate--inhibit, upstream--downstream). & Adds a step that does not belong: irrelevant (L1) or plausible but fabricated and false (L2). \\
    \textbf{What it tests} & Entity grounding and role appropriateness under pathway or system constraints. & Causal semantics and order or sign consistency. & Step existence and mechanistic relevance. \\
    \textbf{Easy (L1)} & Obvious type or species mismatch; simple enzyme swap to a wrong actor. & Textbook flip or subject--object swap; direct polarity inversion. & Clearly off-path module or assay artefact. \\
    \textbf{Hard (L2)} & Paralog, isoform or complex–subunit swap; omission of required PTM or state gating. & Invert upstream--downstream within a complex; alter effect via a single wrong modifier. & Plausible but unsupported step using pathway entities; contradicts curated constraints. \\
    \textbf{Constraints} & Change one entity only; keep verb and polarity identical. & Keep entities identical; only verb changes. & Mechanistic only (no assays). \\
    \bottomrule
  \end{tabularx}
\end{table}

\clearpage
\section{LLM Prompts}
\label{sec:suppmethods-moves-prompts}

\subsection{Game Master}
The game master is a two-step process: \textbf{Diagnose} and \textbf{Move selection}, where the former analyses the current hypothesis and informs the move selection process.

\prompt{supplementary/prompts/diagnose.txt}[diagnose][ChatGPT4o]
\prompt{supplementary/prompts/move_selection.txt}[move-selection][ChatGPT4o]

\subsection{Expanding using LLMs or Corpus}
Expanding a hypothesis consists of two steps: retrieving evidence or information relevant to expansion, and then applying the expansion on the current hypothesis. We provide two ways of retrieving information: (1) via a corpus, and (2) via LLM ``speculation.''

\prompt{supplementary/prompts/retrieve_evidence.txt}[retrieve-evidence][GPT4o]
\prompt{supplementary/prompts/speculate_evidence.txt}[speculate-evidence][GPT4o]
\prompt{supplementary/prompts/expand.txt}[expand][ChatGPT4o]

\subsection{Prune}
\prompt{supplementary/prompts/prune.txt}[prune][ChatGPT4o]

\subsection{Debate — Clash of Claims}
The \textbf{Debate} move is made up of multiple steps.
\begin{itemize}
    \item \textbf{Setup}: An agent that sets up the debate by identifying the key components to be debated, based on the Game Master's request.
    \item \textbf{ClashOfClaims}: A discussion among multiple agents (ClaimSmiths), each starting with a different position on the item being debated.
    \item \textbf{Conclude}: An agent that reads the debate and determines the final conclusion.
\end{itemize}

\prompt{supplementary/prompts/clash_of_claims_setup.txt}[debate-setup][ChatGPT4o]
\prompt{supplementary/prompts/clash_of_claims_conclude.txt}[debate-conclude][ChatGPT4o]
\prompt{supplementary/prompts/claimsmith.txt}[claimsmiths][ChatGPT4o]

\subsection{Baselines}
\prompt{supplementary/prompts/react.txt}[react][ChatGPT4o]
\prompt{supplementary/prompts/chain_of_thought.txt}[chain-of-thought][ChatGPT4o]
\prompt{supplementary/prompts/zero_shot.txt}[zero-shot][ChatGPT4o]

\subsection{User Task Prompts}
\prompt{supplementary/prompts/user_prompt_reconstruction.txt}[reconstruction][ChatGPT4o]
\prompt{supplementary/prompts/user_prompt_corruption.txt}[corruption][ChatGPT4o]

\subsection{Evaluations using LLM-As-Judge}

\subsubsection{Corruption LLM-as-Judge Prompt}

\prompt{supplementary/prompts/llm_as_judge_error_removal.txt}[Error Removal LLM-as-judge][ChatGPT4o]

\subsubsection{Reconstruction}
\prompt{supplementary/prompts/llm_judge_pathway_recall.txt}[Pathway Recall LLM-as-judge][ChatGPT4o]

\subsection{Evaluating LLM-as-judge}
\label{supp-llm-as-judge}

We evaluate the LLM as judge approach against expert human evaluation.
All annotations were collected under strictly blinded, independent conditions: human raters had no access to one another’s labels or to the LLM’s judgments at any point in the labeling process. The human raters were given the exact prompt supplied to the LLM to perform the task.
To obtain a conservative human reference, we define a strict consensus label
\[
\text{human\_consensus} = \mathbbm{1}\{\text{annotator\_1} = 1 \land \text{annotator\_2} = 1\},
\]
i.e., an error is counted as present only when both humans independently flag it.

We quantify agreement using Krippendorff's $\alpha$ for nominal data, computed on the rater-by-item label matrix (\texttt{level\_of\_measurement = "nominal"}).
We adopt Krippendorff's $\alpha$ because it naturally extends to more than two raters and has been recommended in LLM-as-a-judge evaluations over $\kappa$-style and correlation-based measures.

We compute $\alpha$ for (i) \textbf{H1 vs H2} (\texttt{annotator\_1}, \texttt{annotator\_2}), (ii) \textbf{H1 vs LLM} and \textbf{H2 vs LLM} (\texttt{annotator\_1} / \texttt{annotator\_2} with \texttt{llm\_as\_judge}), (iii) \textbf{Consensus vs LLM} (\texttt{human\_consensus}, \texttt{llm\_as\_judge}), and (iv) \textbf{All Raters} (\texttt{annotator\_1}, \texttt{annotator\_2}, \texttt{llm\_as\_judge})

\subsubsection{Corruption LLM-as-Judge Evaluation}
We evaluate the LLM-as-judge of corruption-removal quality in an error-removal task. The judge receives (i) the original (correct) statement, (ii) the corrupted statement, and (iii) the model’s repaired pathway, and must decide whether the original corruption is still present in the repaired pathway.
For each item (a corrupted pathway), the task is binary: label $1$ if a corruption error is still present and $0$ if it has been successfully removed. Two human annotators, both biological experts (\texttt{annotator\_1}, \texttt{annotator\_2}), and the LLM (\texttt{llm\_as\_judge}) independently assign binary labels to each item.
We stratified the dataset by error type (wrong entity, wrong relationship, unsupported statement), difficulty (easy, hard), and source pathway, and sampled uniformly from the resulting strata. Because each annotation required carefully reading the full model output (i.e., the entire pathway) to determine whether an error remained, particularly for long generations, and given the annotators’ time budget, we restricted the evaluation to 20 diverse instances from the full corruptions dataset.
The resulting inter-annotator agreements are summarized in Table~\ref{tab:krippendorff}.

\begin{table}[t]
\centering
\begin{tabular}{l c}
\hline
\textbf{Comparison} & \textbf{Krippendorff's $\alpha$} \\
\hline
H1 vs H2           & 0.900 \\
H1 vs LLM          & 0.900 \\
H2 vs LLM          & 1.000 \\
Consensus vs LLM   & 0.900 \\
All Raters         & 0.933 \\
\hline
\end{tabular}
\caption{Inter-rater agreement (Krippendorff's $\alpha$) between human annotators and the LLM judge on 20 corruption--recovery instances.}
\label{tab:krippendorff}
\end{table}

Although they are based on a small 20-item subset and should therefore be viewed as a sanity check on alignment rather than a precise reliability estimate, these $\alpha$ values ($\geq 0.9$ in all comparisons) nonetheless provide strong evidence of very high agreement between the two human annotators and the LLM judge.

\subsubsection{Reconstruction LLM-as-Judge Evaluation}

We evaluate the LLM-as-judge of the pathway-step reconstruction task. The judge receives (i) the final hypothesis and (ii) the biochemical reaction within a Reactome pathway. If the biochemical reaction is represented in the  hypothesis, the task is binary classification across 4 individual criteria per biochemical reaction: (i) are the correct input entities in the hypothesis (ii) are the correct output entities in the hypothesis (iii) is the directionality of the reaction correct and (iv) is the relation type of the reaction correct. In the case where the reaction is not represented in the text, all labels are to be assigned ``0``. We randomly sampled 5 pathways and all of their constituent biochemical reactions for labelling by 2 human annotators.  The resulting inter-annotator agreements are summarized in Table~\ref{tab:krippendorff-reconstruction}.

\begin{table}[H]
\centering
\begin{tabular}{l l r}
\toprule
\textbf{Annotation Task} & \textbf{Comparison} & \textbf{Krippendorff's $\alpha$} \\
\midrule
input\_entities    & H1 vs H2         & 0.93 \\
                   & H1 vs LLM        & 0.41 \\
                   & H2 vs LLM        & 0.49 \\
                   & Consensus vs LLM & 0.41 \\
                   & All Raters       & 0.61 \\
\midrule
output\_entities   & H1 vs H2         & 0.63 \\
                   & H1 vs LLM        & 0.35 \\
                   & H2 vs LLM        & 0.56 \\
                   & Consensus vs LLM & 0.49 \\
                   & All Raters       & 0.51 \\
\midrule
directionality     & H1 vs H2         & 0.85 \\
                   & H1 vs LLM        & 0.49 \\
                   & H2 vs LLM        & 0.49 \\
                   & Consensus vs LLM & 0.49 \\
                   & All Raters       & 0.61 \\
\midrule
reaction\_type     & H1 vs H2         & 0.76 \\
                   & H1 vs LLM        & 0.70 \\
                   & H2 vs LLM        & 0.62 \\
                   & Consensus vs LLM & 0.69 \\
                   & All Raters       & 0.69 \\
\bottomrule
\end{tabular}
\caption{Inter-rater agreement (Krippendorff's $\alpha$) between human annotators and the LLM judge on the 4 annotation tasks for 5 pathways, totalling 27 distinct pathway steps.}
\label{tab:krippendorff-reconstruction}
\end{table}

Inter annotator agreement is lower than in the simpler corruption annotations. In particular, agreement was poorest between human annotators and the LLM-as-judge on the input and output entity labelling task. After reviewing the annotations with the human annotators, it was found that the variation in labelling stemmed from two sources. Firstly, transient complexes are referenced in Reactome biochemical reactions. Human annotators were stricter than the LLM-as-judge in these annotations, considering explicit reference to complex subunits (rather than the whole transient complex) interacting as insufficiently explicit. Secondly,  Reactome referenced interactions relating large functional families and the hypothesis considered specific family members present in the corpus. In this context, the LLM-as-judge was stricter and considered specific family members to be incorrect entities, whereas human annotators considered specific exemplars as representative of general, well-established signalling pathways.

\section{Reconstruction}

\subsection{Example Reconstruction Game}
\label{sec:suppmethods-reconstruction-game-example}

Below are fragments of text that are added to the hypothesis over the trajectory of a hypothesis expansion game.

\FloatBarrier
\modelOutput{supplementary/modeloutput/reconstruction_game_output.txt}[Translocation of nuclear-encoded proteins into mitochondria.]
\FloatBarrier 

Each addition is granular, and has been informed by a retrieval from the relevant corpus.

\subsection{Reconstruction Additional Results}

\begin{figure}[ht]
  \centering
  \safeincludegraphics[width=0.5\textwidth]{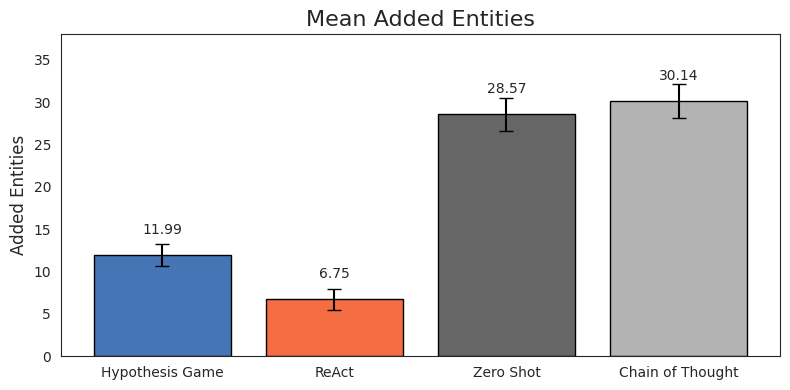}
  \caption{Mean number of added entities with 95\% confidence interval across the 100 pathways from the reconstruction experiments. Added entities are defined as entities (genes, protein complexes/families, and chemicals) not present in the original pathway. \textit{Zero-shot} and \textit{Chain-of-thought} tend to produce long hypotheses with lots of added entities which results in higher recall, but low precision. On the other hand \textit{ReAct} adds less entities which results in higher precision, but low recall. Our method \textit{Hypothesis Game} better balances recall and precision.
} 
  \label{fig:added-entities}
\end{figure}

\subsection{Ablations}
\label{supp:sec:ablations}

\subsubsection{Ablation Design}
To understand how the implemented moves influence the pathway constructions, we ran experiments with various game configurations on $20$ Reactome pathways (distinct from the $100$ used in the main results). The only difference between the game variants was the moves available to the Game Master, other than that all other configurations were the same. 

The ablation results are shown in table~\ref{tab:ablation-combined}. The \textit{Hypothesis Game} uses all $4$ moves, while other game variants are named after the moves they had available. In the current implementation, only the expand move supports retrieval from a corpus, all other moves are based on the LLM's internal knowledge.  To reflect this distinction we categorised the ablation experiments into two categories: 1, \textit{Games using Corpus} where the \texttt{Expand with Corpus} move was available and 2, \textit{Games not using Corpus}. For baselines we used \textit{Zero-shot}, \textit{Chain-of-Thought}, \textit{ReAct} and \textit{ReAct no corpus} (same template as ReAct but without access to the corpus). 

\subsubsection{Ablation Results}

In general, we found the game variants with access to corpus to perform similarly to each other. The \textit{Hypothesis Game} (using all available moves) is marginally better than the other game configurations (precision and F1 scores). Games with retrieval tend to result in slightly better performance across all metrics. Interestingly, \textit{ReAct no corpus} had a much bigger drop-off compared to \textit{ReAct (with corpus)} than observed with the game variants which reinforces the benefits of the available corpus. Even though the games have access to different moves the game master was the one responsible for selecting appropriate moves depending on the current hypothesis state. Since the objective of the reconstruction is to expand an initial hypothesis most of the selected moves were some form of expansion (based on the corpus or LLM knowledge). Overall, the \textit{Hypothesis Game} having access to all moves has shown the benefits of using the moves appropriately to reconstruct the pathways.

\begin{table}[h]
\centering
\begin{tabular}{|l|ccc|}
\hline
\textbf{Method} & \textbf{Recall} & \textbf{Precision} & \textbf{F1 Score} \\
\hline

\multicolumn{4}{|l|}{\textbf{Games using Corpus}} \\
\hline
Hypothesis Game & 0.46 $\pm$ 0.05 & 0.26 $\pm$ 0.03 & 0.31 $\pm$ 0.04 \\
expand\_debate\_prune & 0.48 $\pm$ 0.06 & 0.23 $\pm$ 0.03 & 0.30 $\pm$ 0.03 \\
expand\_debate        & 0.46 $\pm$ 0.06 & 0.23 $\pm$ 0.03 & 0.29 $\pm$ 0.04 \\
expand                & 0.48 $\pm$ 0.06 & 0.26 $\pm$ 0.05 & 0.30 $\pm$ 0.04 \\
\hline

\multicolumn{4}{|l|}{\textbf{Games not using Corpus}} \\
\hline
expand\_debate\_prune  & 0.43 $\pm$ 0.06 & 0.24 $\pm$ 0.04 & 0.28 $\pm$ 0.04 \\
expand\_debate         & 0.37 $\pm$ 0.05 & 0.22 $\pm$ 0.03 & 0.25 $\pm$ 0.04 \\
expand                 & 0.44 $\pm$ 0.06 & 0.21 $\pm$ 0.03 & 0.27 $\pm$ 0.03 \\
debate                 & 0.39 $\pm$ 0.06 & 0.17 $\pm$ 0.03 & 0.21 $\pm$ 0.02 \\
\hline

\multicolumn{4}{|l|}{\textbf{Baselines}} \\
\hline
ReAct (corpus) & 0.40 $\pm$ 0.06 & 0.35 $\pm$ 0.05 & 0.32 $\pm$ 0.05 \\
ReAct no corpus    & 0.40 $\pm$ 0.06 & 0.26 $\pm$ 0.05 & 0.25 $\pm$ 0.03 \\
Zero-shot        & 0.56 $\pm$ 0.05 & 0.14 $\pm$ 0.02 & 0.22 $\pm$ 0.02 \\
Chain-of-Thought & 0.58 $\pm$ 0.06 & 0.15 $\pm$ 0.02 & 0.22 $\pm$ 0.03 \\
\hline

\end{tabular}
\caption{Comparison of different game variants vs. prompting baselines on 20 additional pathway construction task. The entries show mean entity-level recall, precision, and F1 scores with standard error, grouped by method family. Note that the games in the section ``Games not using Corpus'' only had access to the \texttt{Expand without Corpus} move, while in the ``Games using Corpus'' only had access to \texttt{Expand with Corpus}, except Hypothesis Game that had access to both types of expand moves.}
\label{tab:ablation-combined}
\end{table}

\section{Corruption}

\subsection{Stratified Corruption Results}
To further probe system behavior in the corruption task in addition to stratifying performance by error type, we also stratify performance by error difficulty and error fraction.
\label{sec:coruption-stratified-results}

\begin{figure}[h]
    \centering
    \safeincludegraphics[width=\textwidth]{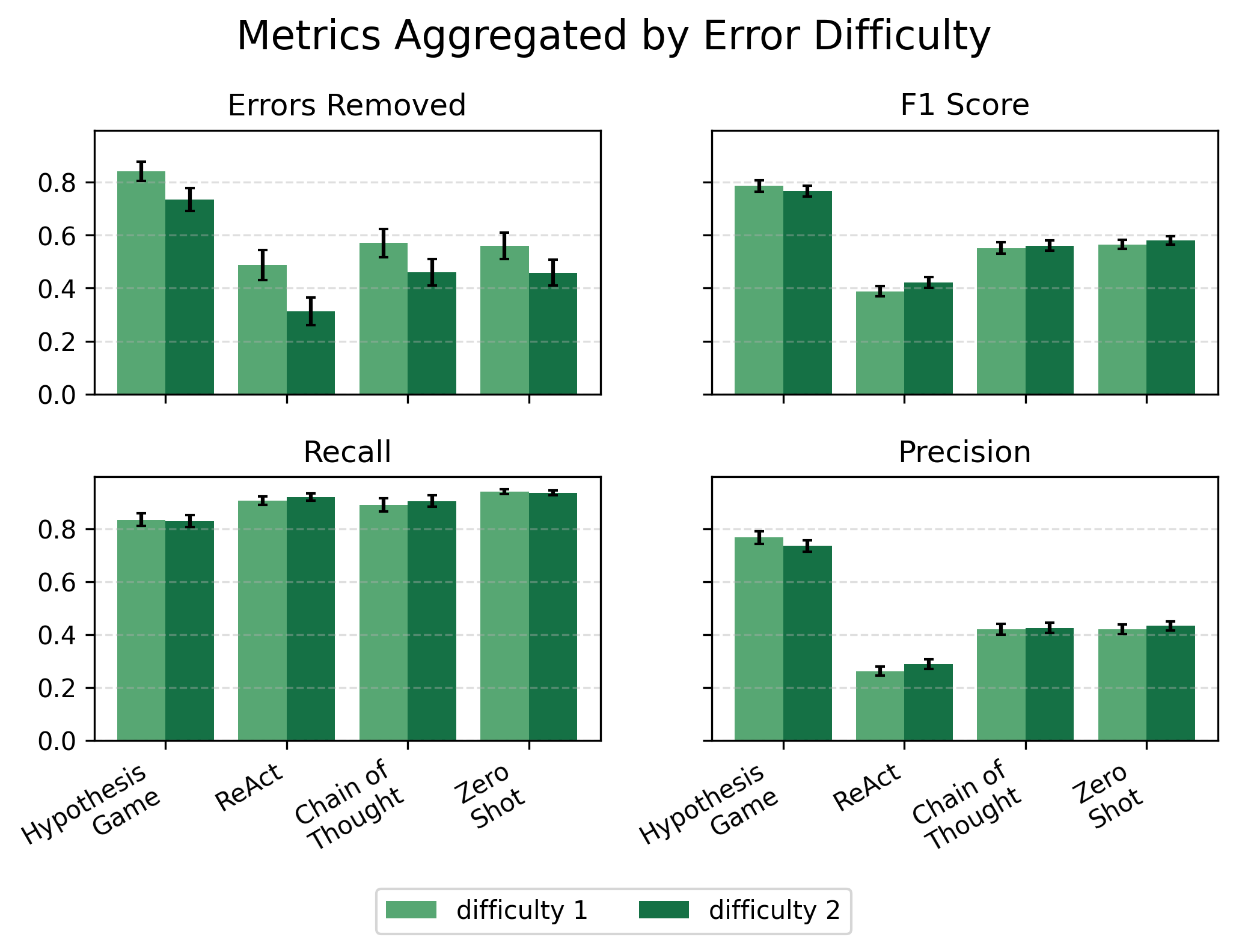}
    \caption{Aggregation of all results on the corruption task based on error difficulty. Error bars show 95\% confidence intervals.}
    \label{fig:error_difficulty}
\end{figure}

\paragraph{Error difficulty.} Figure~\ref{fig:error_difficulty} confirms the expected separation between easier (L1) and harder (L2) variants. Harder corruptions have lower error removal rates across all models. Interestingly, recall and precision remain relatively stable across difficulty levels, indicating that difficulty primarily affects the detectability of corrupted statements rather than the fidelity of pathway reconstruction once errors are removed.

\begin{figure}[h]
    \centering
    \safeincludegraphics[width=\textwidth]{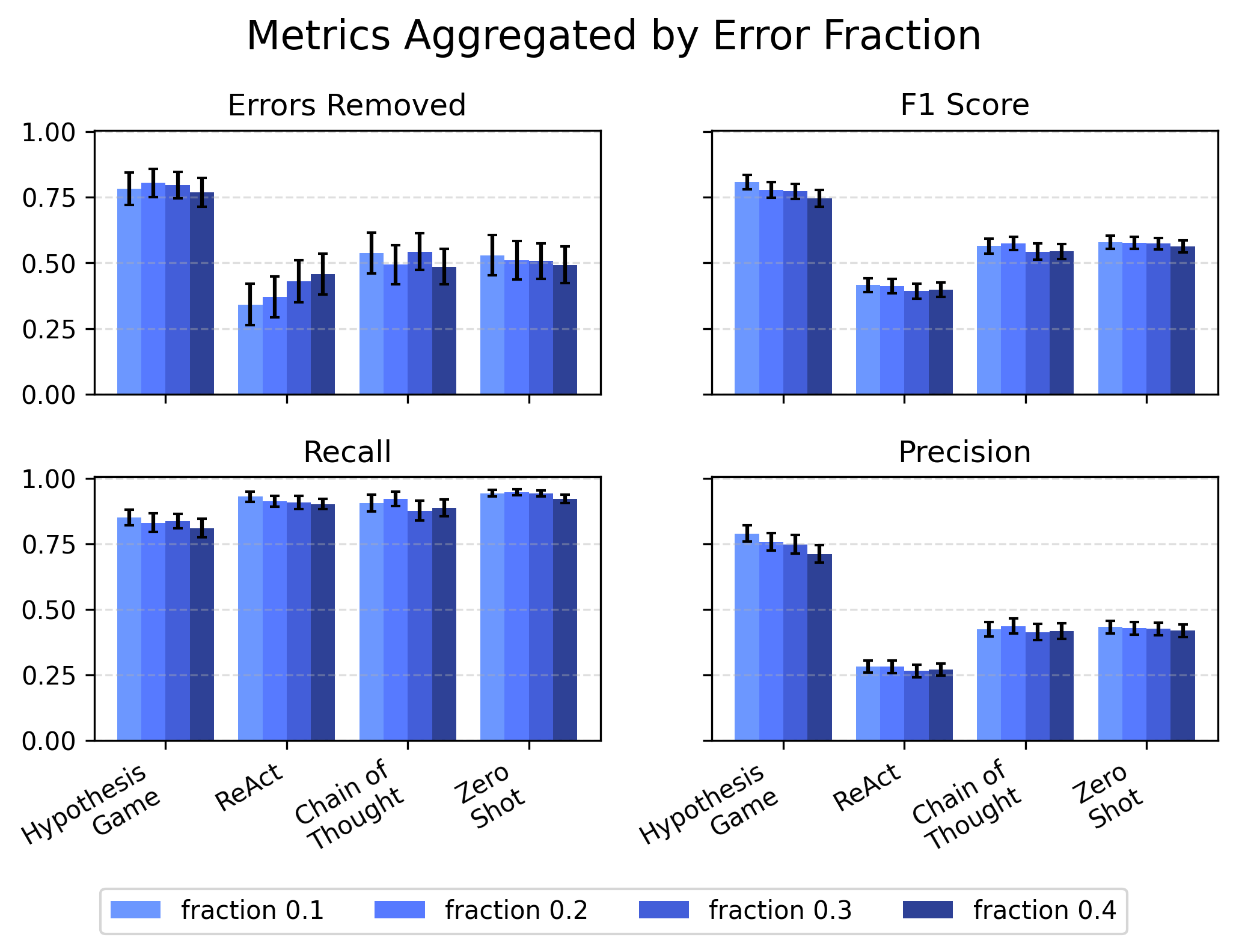}
    \caption{Aggregation of all results on the corruption task based on error fraction. Error bars show 95\% confidence intervals.}
    \label{fig:error_frac}
\end{figure}

\paragraph{Error fraction.} Finally, Figure~\ref{fig:error_frac} examines robustness to increasing corruption density. Performance is remarkably stable across fractions: even when 40\% of pathway steps are corrupted, removal, recall, and precision degrade only mildly. This suggests that model strategies scale linearly rather than collapsing under higher noise levels, pointing to robustness at the pathway level rather than brittleness to compounded errors. In future work we plan to investigate the effect of increasing the percentage of errors beyond 40\%.


\paragraph{} Overall, these stratified analyses show that error type and difficulty shape the challenge in meaningful ways, while corruption density has a surprisingly limited impact.

\subsection{Extent of Hypothesis Modification}
\label{sec:hypothesis-modification}

\begin{figure}[H]
    \centering
    \safeincludegraphics[width=\textwidth]{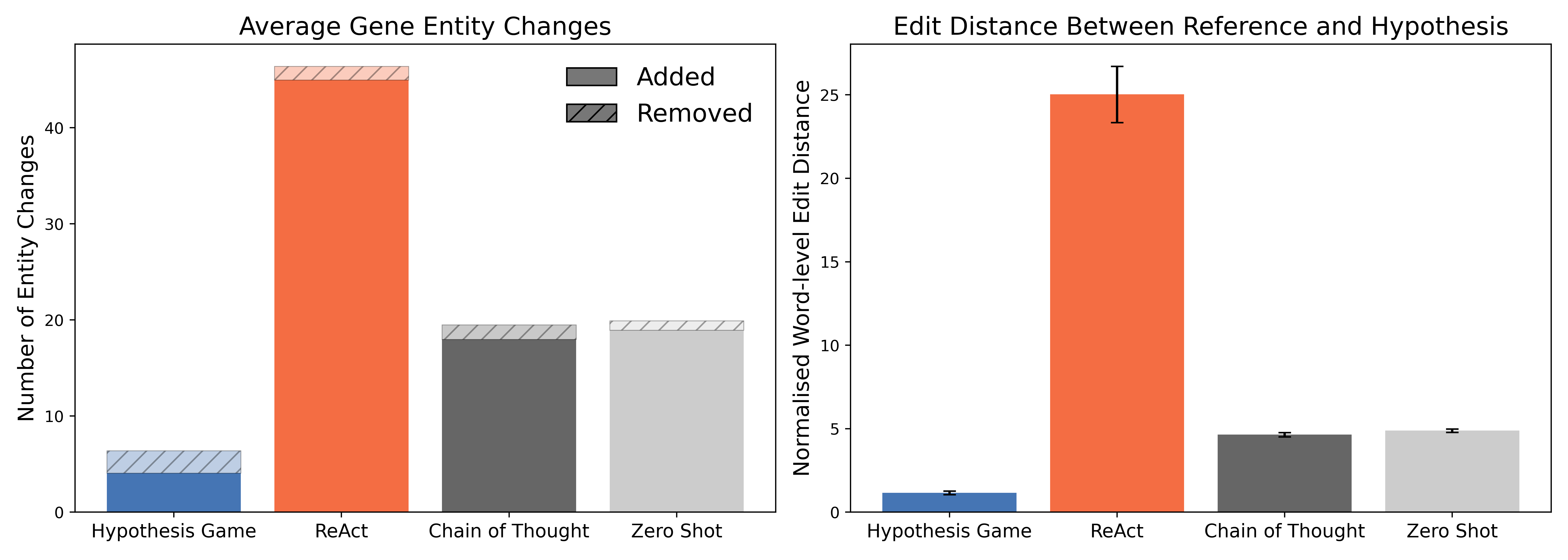}
    \caption{Extent of hypothesis modification across models. \textbf{Left:} Number of gene-level entity changes (additions or removals) identified using Gilda. \textbf{Right:} Word-level normalised Levenshtein distance to the reference pathway description. Error bars show 95\% confidence intervals.}
    \label{fig:hypothesis_drift}
\end{figure}

To assess how much each reasoning model alters the original pathway description during refinement, in Figure~\ref{fig:hypothesis_drift} we quantify differences between the model's final output and the ground-truth Reactome reference. This serves as a sanity check for over-editing and complements our corruption evaluation by revealing how much the models deviate from an error-free reference.

\paragraph{Entity-level changes.}  
We compute the total number of gene-level entities that are either added or removed during hypothesis refinement. Entities are identified using \texttt{Gilda}-tagged named entity recognition, consistent with the rest of our evaluation pipeline. This metric captures biologically meaningful modifications to the pathway hypothesis. A higher value indicates greater divergence from the reference, either due to correction or unnecessary hallucination. We report the mean entity change count per model, with 95\% confidence intervals.

\paragraph{Text-level changes.}  
To complement entity-level analysis, we also compute the \emph{word-level normalised Levenshtein distance} between the final hypothesis and the reference. This metric measures the minimal number of word insertions, deletions, or substitutions required to transform the reference into the model's output, normalised by the reference word count. Unlike the entity metric, this captures broader forms of rewriting such as paraphrasing and reordering, regardless of biological content.

\paragraph{Interpretation.}  
Figure~\ref{fig:hypothesis_drift} shows that models using explicit planning strategies, such as Hypothesis Game, make fewer changes at both the semantic (entity) and surface (text) levels. ReAct, in contrast, tends to revise more aggressively. Importantly, we observe aligned trends across both metrics—entity changes and text distance—suggesting robustness of the conclusion across both biologically grounded and lexical measures.

\subsection{Example Final Hypothesis}
\ref{tab:corruption_example_table} compares example final hypothesis from Hypothesis Game and ReAct. The example was computed on the Reactome pathway R-HSA-1268020, with the following corruption policy:  

\begin{quote}
\begin{itemize}
    \item \textbf{Error type:} wrong entity
    \item \textbf{Error difficulty:} 2
    \item \textbf{Error fraction:} 0.3 (4 errors)
\end{itemize}
\end{quote}

The errors introduced are shown in \ref{tab:corruption_example_statements}.
\begin{table}[H]
\centering
\begin{tabular}{c p{0.42\textwidth} p{0.42\textwidth}}
\toprule
\# & \textbf{Original statement} & \textbf{Corrupted statement} \\
\midrule
1 & TOMM40 complex translocates proteins from the cytosol to the mitochondrial intermembrane space 
  & Mitochondrial intermembrane space translocates proteins into the cytosol via TOMM40 complex \\
\midrule
2 & MIA40:ERV1 (CHCHD4:GFER) oxidizes cysteine residues to cystine disulfide bonds 
  & Cystine disulfide bonds oxidize MIA40:ERV1 (CHCHD4:GFER) \\
\midrule
3 & MPP cleaves targeting peptide (presequence) of inner membrane precursors 
  & MPP ligates targeting peptide to inner membrane precursors \\
\midrule
4 & PITRM1 proteolyzes mitochondrial targeting peptides (presequences) 
  & PITRM1 stabilizes mitochondrial targeting peptides (presequences) \\
\bottomrule
\end{tabular}
\caption{Examples of original statements and statements corrupted with wrong direction errors introduced into \ref{tab:corruption_example_table}.}
\label{tab:corruption_example_statements}
\end{table}

\renewcommand{\arraystretch}{1.3} 

\begin{longtable}{%
  p{0.27\linewidth} 
  p{0.27\linewidth} 
  p{0.4\linewidth}  
}
\caption{Comparison of Hypothesis Game and ReAct on the Mitochondrial Import Pathway.}
\label{tab:corruption_example_table} \\
\toprule
\textbf{Reactome Reference} & \textbf{Hypothesis Game} & \textbf{ReAct} \\
\midrule
\endfirsthead

\toprule
\textbf{Reactome Reference} & \textbf{Hypothesis Game} & \textbf{ReAct} \\
\midrule
\endhead

Mitochondrial protein import &
Pathway Name: mitochondrial protein import &
Pathway Name: Mitochondrial Protein Import \\
\hline

 &  &
Begins with the targeting of cytosol-synthesized proteins to mitochondria via specific signals. \\
\hline

TOMM40 complex translocates proteins from the cytosol to the mitochondrial intermembrane space &
Step 1. TOMM40 imports proteins from the cytosol into the mitochondrial intermembrane space &
The TOM complex, including Tom40, serves as the initial entry gate for precursor proteins, crucial for metabolite exchange and apoptosis regulation. \\
\hline

MIA40:ERV1 (CHCHD4:GFER) oxidizes cysteine residues to cystine disulfide bonds &
Step 2. MIA40 oxidizes incoming proteins, and ERV1 reoxidizes MIA40 &
The disulfide relay system (DRS) in the intermembrane space (IMS), involving Mia40 and Erv1 (GFER in humans), facilitates the import of cysteine-rich proteins via oxidative folding, linked to the respiratory chain. \\
\hline

TIMM8:TIMM13 chaperones hydrophobic proteins &
Step 3. TIMM8:TIMM13 chaperones hydrophobic proteins &
TIMM8a and TIMM13 form a complex in the IMS to import inner membrane substrates like TIMM23, while the TIM22 complex, involving Tim9, Tim10a, and Tim10b, inserts carrier preproteins into the inner membrane. \\
\hline

TIMM9:TIMM10 binds hydrophobic proteins &
Step 4. TIMM9:TIMM10 binds hydrophobic proteins &
\\
\hline

TIMM9:TIMM10 transfers proteins to TIMM22 &
Step 5. TIMM9:TIMM10 transfers proteins to TIMM22 &
\\
\hline

SAM50 complex inserts proteins into mitochondrial outer membrane &
Step 6. SAM50 inserts $\beta$-barrel proteins &
Sam50, part of the sorting and assembly machinery (SAM), works with the TOM complex for the import and assembly of outer membrane proteins like VDAC. \\
\hline

TIMM22 inserts proteins into inner membrane &
Step 7. TIMM22 inserts proteins into inner membrane &
\\
\hline

Precursor proteins enter TIMM23 SORT &
Step 8. Precursor proteins enter TIMM23 for inner membrane insertion &
The TIM23 complex, the main entry gate for matrix and inner membrane proteins, forms the import pore across the inner membrane, interacting electrostatically with presequence peptides. \\
\hline

 &
Step 8a. The membrane potential ($\Delta \psi$) across the inner mitochondrial membrane provides the driving force for the initial translocation of precursor proteins through the TIMM23 complex. &
\\
\hline

MPP cleaves targeting peptide (presequence) of inner membrane precursors &
Step 9. MPP cleaves targeting peptide to inner membrane precursors &
\\
\hline

TIMM23 SORT inserts proteins into inner membrane &
Step 10. TIMM23 inserts proteins into inner membrane &
\\
\hline

Precursor proteins enter TIMM23 PAM &
Step 11. Precursor proteins enter TIMM23 PAM &
\\
\hline

 &  &
Mitochondrial Hsp70 (mtHsp70) unfolds precursor proteins to facilitate translocation, working with the TIM23 complex. \\
\hline

MPP cleaves targeting peptide (presequence) of matrix precursors &
Step 12. MPP cleaves targeting peptide (presequence) of matrix precursors &
Mitochondrial processing peptidase (MPP) cleaves targeting peptides of matrix precursors, while PITRM1 stabilizes mitochondrial targeting peptides (presequences) and degrades amyloid beta-protein (Abeta). \\
\hline

TIMM23 PAM translocates proteins from the mitochondrial intermembrane space to the mitochondrial matrix &
Step 13. TIMM23 PAM translocates proteins from the mitochondrial intermembrane space to the mitochondrial matrix &
\\
\hline

PITRM1 proteolyzes mitochondrial targeting peptides (presequences) &
Step 14. PITRM1 degrades presequences &
\\
\hline

 &  &
Feedback and compensatory mechanisms include redox regulation by conserved cysteine residues, prevention of precursor protein aggregation by receptor domains like Tom70, and integration of protein import with mitochondrial energetics through the disulfide relay system's link to the respiratory chain. \\
\bottomrule
\end{longtable}


\bibliographystyle{plainnat}
\bibliography{references}
\end{document}